\pdfoutput=1
\documentclass[12pt]{article}
\usepackage[margin=1in]{geometry}
\usepackage{amsmath}
\usepackage{titling}
\usepackage{blindtext}
\usepackage{pgfplots}
\usepackage{natbib}
\usepackage{comment}
\usepackage[colorlinks=true,linkcolor=blue,allcolors=blue]{hyperref}
\usepackage{url}
\usepackage{setspace}
\usepackage{authblk}
\pgfplotsset{compat=1.14}

\providecommand{\keywords}[1]
{
  \small	
  \textit{Keywords:} #1
}

\begin{document}

\title{Effects of Cowling Resistivity in the Weakly-Ionized Chromosphere}


\author[1]{M. S. Yalim}
\affil[1]{Center for Space Plasma and Aeronomic Research, The University of Alabama in Huntsville, Huntsville, AL 35805, USA}

\author[1]{A. Prasad}

\author[1,2]{N. V. Pogorelov}
\affil[2]{Department of Space Science, The University of Alabama in Huntsville, Huntsville, AL 35805, USA}

\author[1,2]{G. P. Zank}

\author[1,2]{Q. Hu}

\setcounter{Maxaffil}{0}
\renewcommand\Affilfont{\itshape\small}
\date{}    
\begin{titlingpage}
    \maketitle
\begin{abstract}
The physics of the solar chromosphere is complex from both theoretical and modeling perspectives. The plasma temperature from the photosphere to corona increases from $\sim$5,000 K to $\sim$1 million K over a distance of only $\sim$10,000 km from the chromosphere and the transition region. Certain regions of the solar atmosphere have sufficiently low temperature and ionization rates to be considered as weakly-ionized. In particular, this is true at the lower chromosphere. As a result, the Cowling resistivity is orders of magnitude greater than the Coulomb resistivity. Ohm's law therefore includes anisotropic dissipation. To evaluate the Cowling resistivity, we need to know the external magnetic field strength and to estimate the neutral fraction as a function of the bulk plasma density and temperature. In this study, we determine the magnetic field topology using the non-force-free field (NFFF) extrapolation technique based on \emph{SDO}/HMI SHARP vector magnetogram data, and the stratified density and temperature profiles from the Maltby-M umbral core model for sunspots. We investigate the variation and effects of Cowling resistivity on heating and magnetic reconnection in the chromosphere as the flare-producing active region (AR) 11166 evolves. In particular, we analyze a C2.0 flare emerging from AR11166 and find a normalized reconnection rate of 0.051.


\end{abstract}
\end{titlingpage}



\section{Introduction} \label{sec:intro}

The lower atmosphere of the Sun (i.e., photosphere and chromosphere) is composed of weakly-ionized plasma. The low temperature in the photosphere results in an ionization fraction of about $n_{i}/n\approx 10^{-4}$ where 1 corresponds to fully ionized plasma. In the chromosphere, the ionization fraction increases but always remains below 1~\citep{Khomenko16}. The dominant mechanism for ionization in the chromosphere is photoionization with the rate for hydrogen $\approx$ 0.014 $\mathrm{s}^{-1}$, which is orders of magnitude greater than the ionization rate due to electron collisions of $\approx 7.8\times10^{-5}$ $\mathrm{s}^{-1}$~\citep{PM98}.

The plasma $\beta$ exceeds 1 in the photosphere except for sunspot locations, and rapidly decreases in the chromosphere~\citep{Gary01}. The plasma $\beta$ being high in the photosphere, typical coronal magnetic field extrapolation techniques based on nonlinear-force-free-fields (NLFFFs), which are widely accepted by the solar community~\citep{WS12}, do not work very well as the Lorentz force is non-negligible in both photosphere and lower chromosphere.
A novel alternative to NLFFF is an extrapolation using non-force-free-fields (NFFFs), which are described by the double-curl Beltrami equation for the magnetic field $\textbf{\emph{B}}$, derived from the variational principle of minimum energy dissipation rate~\citep{Bhattacharyya07}. The equation was first solved analytically to obtain magnetic flux ropes that resembled coronal loops~\citep{Bhattacharyya07}. This technique is discussed in more detail in section~\ref{cowling}.

\citet{Cowling57} showed that the electrical conductivity (the Cowling conductivity) of a non-stationary plasma can be significantly decreased owing to ion acceleration by Ampere's force. Collisions between ion and neutral particles become very effective because of the high ion velocities. As a result, the magnetic flux is not conserved and the rate of magnetic reconnection might be considerably increased due to Joule's (Ohmic) dissipation~\citep{TS10}.

From the perspective of energy balance in the weakly-ionized chromosphere, Cowling resistivity leads to additional dissipation of currents perpendicular to the magnetic field resulting in Joule heating that is several orders of magnitude larger compared to the fully ionized plasma.

Section~\ref{cowling} describes the calculation of the Cowling resistivity. Section~\ref{res} presents results and discusses the effects of Cowling resistivity on heating and magnetic reconnection in the chromosphere. In particular, we follow the evolution of AR11166, and its effect on magnetic reconnection and the formation of a C2.0 flare. Finally, section~\ref{conc} presents our conclusions.     

\section{Calculation of the Cowling Resistivity} \label{cowling}

To describe the interaction of chromospheric plasma with magnetic field, and its dependence on the degree of collisional coupling, we may apply a quasi-MHD single fluid theory complemented with a generalized Ohm's law, or we may treat neutral and charged fluids separately as fluids interacting by collisions \citep[e.g.][]{Leake12}. In this paper, we adopt the former approach according to the formulations given in~\citet{LA06}.

Our main focus is on the calculation and effects of Cowling resistivity. The anisotropic dissipation of currents due to the presence of Cowling resistivity can be seen in the induction and energy equations given in Eq.~\ref{eqChrom1a} and Eq.~\ref{eqChrom1b}, respectively that illustrate the impact of the presence of neutrals through the terms that contain $\eta_{C}$:
\begin{equation}
\label{eqChrom1a}
\frac{\partial\textbf{\emph{B}}}{\partial t}+\mathbf{\nabla}\cdot(\textbf{\emph{v}}\textbf{\emph{B}}-\textbf{\emph{B}}\textbf{\emph{v}})+\mathbf{\nabla}\times\eta\textbf{$\emph{J}_{\parallel}$}+\mathbf{\nabla}\times\eta_{C}\textbf{$\emph{J}_{\perp}$}=0,
\end{equation}
and
\begin{equation}
\label{eqChrom1b}
\frac{\partial E}{\partial t}+\mathbf{\nabla}\cdot\Big[(E+p+\frac{B^2}{8\pi})\textbf{\emph{v}}-\frac{\textbf{\emph{B}}}{4\pi}(\textbf{\emph{v}}\cdot\textbf{\emph{B}})\Big]-\mathbf{\nabla}\cdot (\textbf{\emph{B}}\times\eta\textbf{$\emph{J}_{\parallel}$})-\mathbf{\nabla}\cdot (\textbf{\emph{B}}\times\eta_{C}\textbf{$\emph{J}_{\perp}$})=\rho(\textbf{\emph{v}}\cdot\textbf{\emph{g}})+S_{NA}.
\end{equation}\noindent
Here, $\eta$ is the Coulomb resistivity, $\eta_{C}$ is the Cowling resistivity, $\textbf{$\emph{J}_{\parallel}$}$ and $\textbf{$\emph{J}_{\perp}$}$ are the components of current density parallel and perpendicular to the magnetic field \textbf{\emph{B}}, $E$, $p$, $\rho$, \textbf{\emph{v}}, and \textbf{\emph{g}} are specific total energy, thermal pressure, density, velocity, and gravitational acceleration, respectively, and $S_{NA}$ is the combination of non-adiabatic source terms corresponding to viscous heating, shock heating, thermal conduction, radiative transfer, and coronal heating.

Accordingly, the Cowling resistivity dissipates currents perpendicular to the magnetic field while the Coulomb resistivity dissipates currents parallel to it. In addition, Cowling resistivity contributes to heating the chromosphere via the frictional Joule heating term that follows from the generalized Ohm's law according to~\citet{LA06}:
\begin{equation}
\label{eqChrom2}
Q=(\textbf{\emph{E}}+(\textbf{\emph{v}}\times\textbf{\emph{B}}))\cdot\textbf{\emph{j}}=\eta J_{\parallel}^2+\eta_{C}J_{\perp}^2.
\end{equation}
To evaluate the expression for the Cowling resistivity, $\eta_{C}$, an estimate for the neutral fraction $\xi_{n}$ is required as a function of density and temperature (to be described below). 

Following the method of~\citet{DePontieu99} an electro-neutral hydrogen plasma is assumed. The solar chromosphere is not in LTE, hence a simple one-level model for the hydrogen atom is inadequate for these conditions~\citep{PT59}. To calculate ionization degrees in a non-LTE situation requires the solution of the radiative transfer and statistical equilibrium equations. These are very time consuming to calculate. For this reason, approximations of non-LTE effects on hydrogen ionization have been developed. Accordingly, a two-level model is used for the hydrogen atom, as this provides us with a good approximation for hydrogen ionization at chromospheric densities and temperatures~\citep{TA61}. Under this approximation, the ionization equation~\citep{Brown73} is solved assuming that thermal collisional ionization is not important when compared to photoionization~\citep{Ambartsumyan58}. The steady state solution to this equation is given by~\citet{TA61} (i.e., the modified Saha equation for non-LTE chromosphere):
\begin{equation}
\label{eqChrom3}
\frac{n_{i}^2}{n_{n}}=\frac{f(T)}{b(T)}, 
\end{equation}
with
\begin{equation}
\label{eqChrom4}
f(T)=\frac{(2\pi m_{e}k_{B}T)^{3/2}}{h^3}\mathrm{exp}\Bigl(-\frac{X_{i}}{k_{B}T}\Bigr), 
\end{equation}
and 
\begin{equation}
\label{eqChrom5}
b(T)=\frac{T}{w T_{R}}\mathrm{exp}\left[\frac{X_{i}}{4 k_{B}T}\left(\frac{T}{T_{R}}-1\right)\right],
\end{equation}
where $k_{B}$ is the Boltzmann constant, $h$ is Planck's constant, $X_{i}$ is the ionization energy of the hydrogen atom, $T_{R}$ is the temperature of the photospheric radiation field and $w$ is its dilution factor.

Using Eq.~\ref{eqChrom3}, the ratio of the number density of neutrals to ions is given by 
\begin{equation}
\label{eqChrom6}
r=\frac{n_{n}}{n_{i}}=\frac{1}{2}\left(-1+\sqrt{\left(1+\frac{4\rho/ m_{i}}{n_{i}^2/ n_{n}}\right)}\right),
\end{equation}
and $\xi_{n}=\frac{\rho_{n}}{\rho}=\frac{r}{1+r}$ is the neutral fraction for a hydrogen plasma (where $m_{i}=m_{n}$). The approximation $\rho\approx m_{i}n_{i}+m_{n}n_{n}=m_{i}(n_{i}+n_{n})$ is used as the mass of the electron is small compared with the proton/neutron.

In this study, we prefer to utilize the steady state solution to the ionization equation via the modified Saha equation to solving non-equilibrium ionization of hydrogen in a time-dependent manner~\citep{Martinez-Sykora20} since we calculate $\eta_{C}$ based on the SHARP vector magnetogram data from the Helioseismic and Magnetic Imager \citep[HMI;][]{Schou12} onboard the Solar Dynamics Observatory \citep[\emph{SDO};][]{Pesnell12} at 13 timesteps with a cadence of 8 hours (see subsection~\ref{evol}).

The relation between the Cowling and Coulomb resistivities is
\begin{equation}
\label{eqChrom7}
\frac{\xi_{n}^2 B_{0}^2}{\alpha_{n}}=\eta_{C}-\eta,
\end{equation}
where $B_{0}$ is the magnetic field strength and $\alpha_{n}=m_{e}n_{e}\nu_{en}^\prime+m_{i}n_{i}\nu_{in}^\prime$ with $\nu_{en}^\prime$ and $\nu_{in}^\prime$ defined as the effective collisional frequencies of electrons and ions with neutrals, respectively. Assuming that the chromospheric plasma is entirely composed of hydrogen, 
\begin{equation}
\label{eqChrom8}
\alpha_{n}=\frac{1}{2}\xi_{n}\big(1-\xi_{n}\big)\frac{\rho^{2}}{m_{n}}\sqrt{\frac{16 k_{B}T}{\pi m_{i}}}\Sigma_{in}, 
\end{equation}
where $\Sigma_{in}$ is the ion-neutral cross-section for a hydrogen plasma.

The Coulomb resistivity is computed from 
\begin{equation}
\label{eqChrom9}
\eta=\frac{m_{e}\big(\nu_{ei}^\prime+\nu_{en}^\prime\big)}{n_{e}e^{2}},
\end{equation}
where $e$ is the charge of an electron, $\nu_{en}^\prime$ and $\nu_{ei}^\prime$ are the effective collisional frequencies of electrons with neutrals and ions given by 
\begin{equation}
\label{eqChrom10}
\nu_{en}^\prime=\frac{m_{n}}{m_{n}+m_{e}}\nu_{en},
\end{equation}
and
\begin{equation}
\label{eqChrom11}
\nu_{ei}^\prime=\frac{m_{i}}{m_{i}+m_{e}}\nu_{ei}.
\end{equation} 
Following the example of~\citet{Spitzer62}, the collisional frequencies of electrons with neutrals and ions are estimated by 
\begin{equation}
\label{eqChrom12}
\nu_{en}=n_{n}\sqrt{\frac{8 k_{B} T}{\pi m_{en}}}\Sigma_{en},
\end{equation}
and 
\begin{equation}
\label{eqChrom13}
\nu_{ei}=3.7\times10^{-6}\frac{n_{i}(\mathrm{ln}\Lambda) Z^{2}}{T^{3/2}}, 
\end{equation}
where $m_{en}=\frac{m_{e} m_{n}}{m_{e}+m_{n}}$, $\Sigma_{en}$ is the electron-neutral cross-section for a hydrogen plasma, $n_{n}=rn_{i}$ is the neutral number density, $Z$ is the atomic number of hydrogen, and $\mathrm{ln}\Lambda$ is the Coulomb logarithm tabulated in~\citet{Spitzer62}.

\begin{figure}[!http]
\centering
\includegraphics[width=0.485\textwidth]{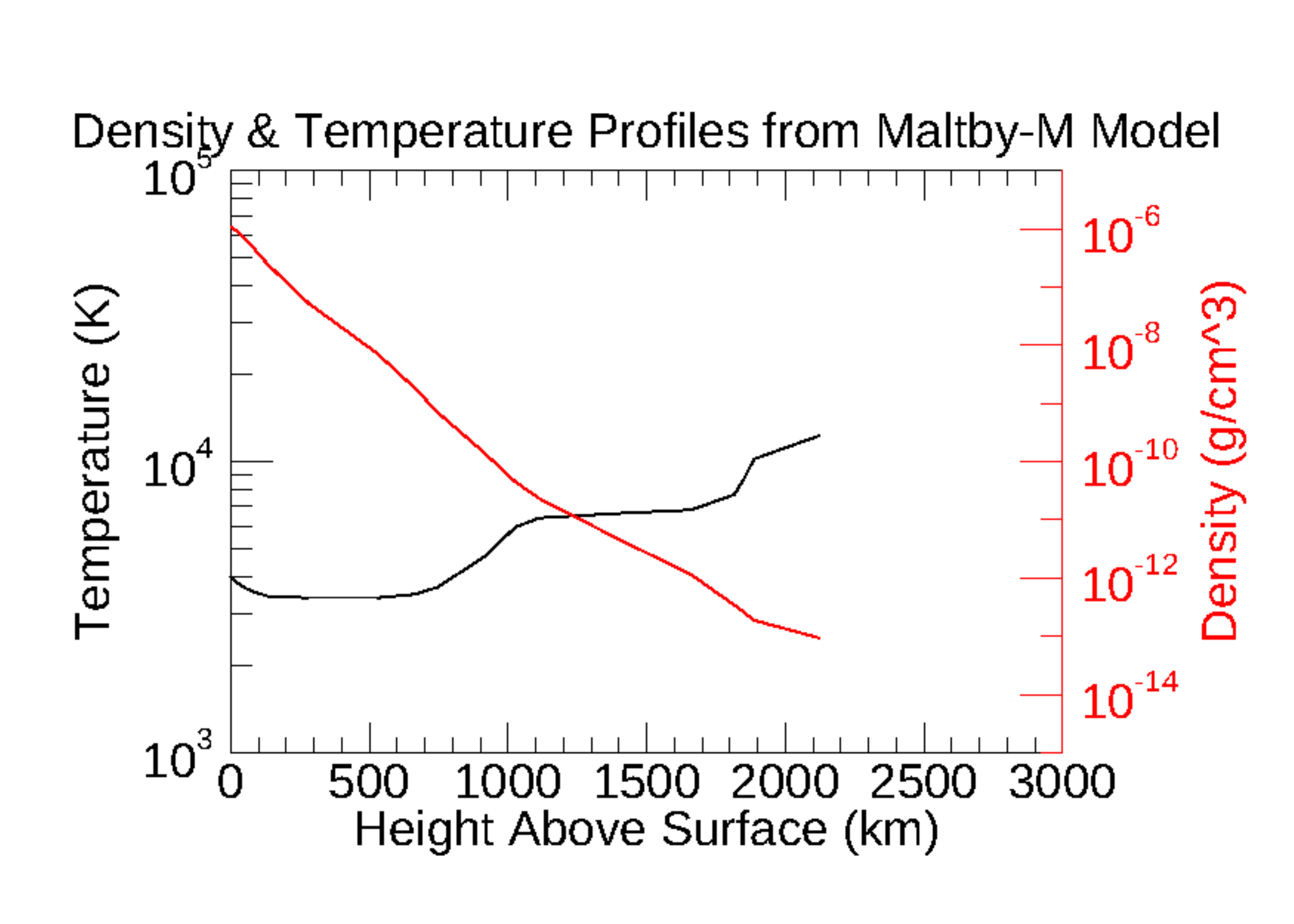}
\includegraphics[width=0.485\textwidth]{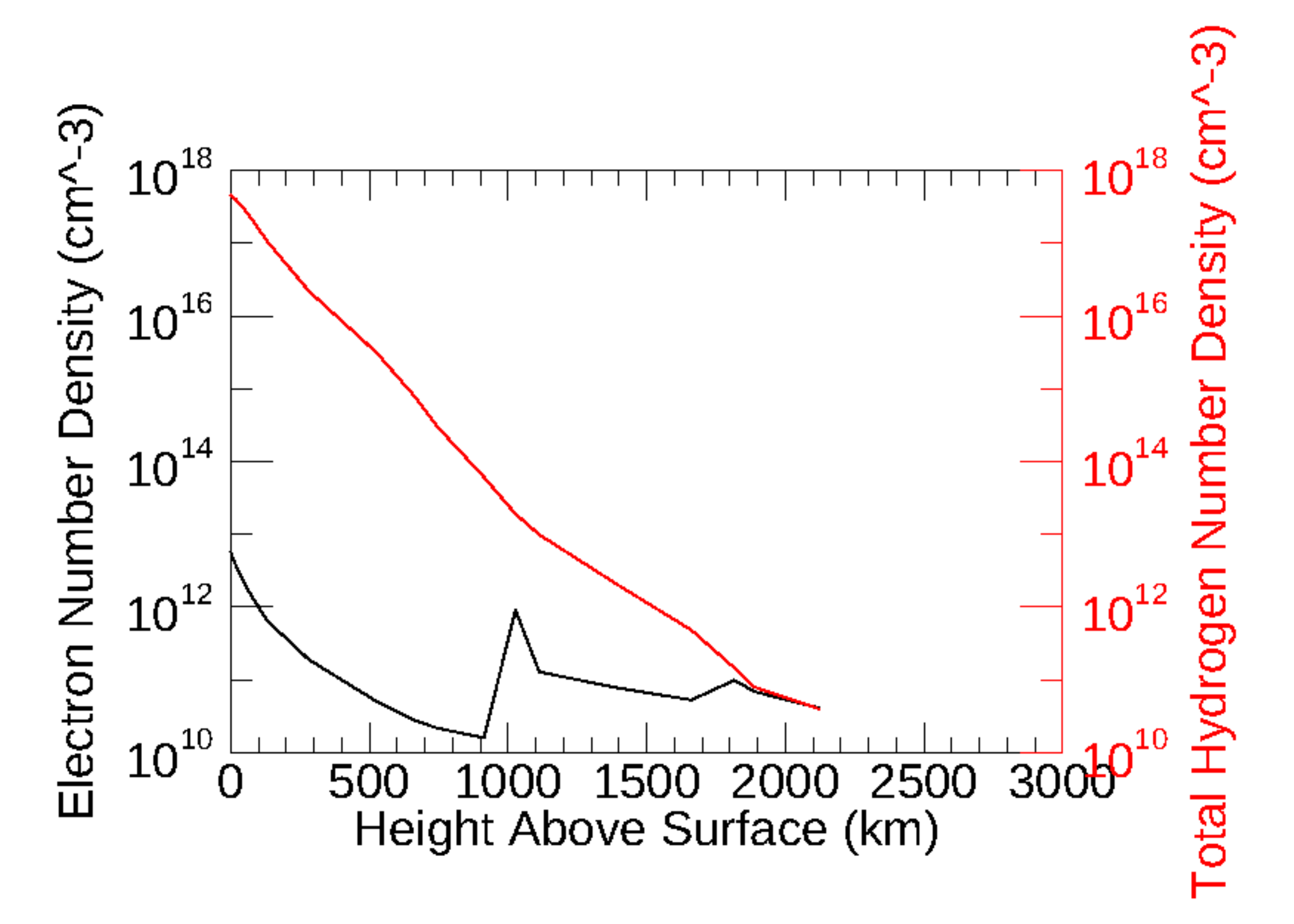}
\caption{(Left) $\rho$ and $T$, (right) $n_{e}$ and $n_{H}$ profiles in the chromosphere obtained from the Maltby-M model.}
\label{fig1}
\end{figure}

To calculate the Coulomb and Cowling resistivities, we need the plasma bulk density $\rho$ and temperature $T$ as well as the ion and electron number densities, $n_{i}$ and $n_{e}$, in the chromosphere where $\rho$, $T$, $n_{e}$, and $n_{H}$, which is the total hydrogen number density, are tabulated by the Maltby-M umbral core model~\citep{Maltby86} for sunspots (see Figure~\ref{fig1}). To calculate $n_{i}$, we use $n_{i}=\frac{n_{H}-n_{e}}{r+1}$. We compute the magnetic field from the NFFF extrapolation technique based on the photospheric vector magnetograms from \emph{SDO}/HMI SHARP data series.

The NFFF extrapolation technique used in this paper was developed by~\citet{HD08,Hu08,Hu10}. Here, the magnetic field $\textbf{\emph{B}}$ is written as
\begin{equation}
\label{eqChrom14}
\textbf{\emph{B}} = \textbf{$\emph{B}_{1}$}+\textbf{$\emph{B}_{2}$}+\textbf{$\emph{B}_{3}$}; \quad \mathbf{\nabla}\times\textbf{$\emph{B}_{i}$}=\alpha_i \textbf{$\emph{B}_{i}$}
\end{equation}
where, for $i=1,2,3$, each subfield $\textbf{$\emph{B}_{i}$}$ corresponds to a linear-force-free field (LFFF) with corresponding constants $\alpha_i$. Further, without loss of generality, we choose $\alpha_1\ne\alpha_3$ and $\alpha_2 = 0$ making $\textbf{$\emph{B}_{2}$}$ a potential field. Subsequently, an optimal pair of $\alpha=\{\alpha_1 , \alpha_3\}$ is obtained by an iterative trial-and-error method which finds the pair that minimizes the average deviation between the observed ($\textbf{$\emph{B}_{t}$}$) and the calculated ($\textbf{$\emph{b}_{t}$}$) transverse fields on the photospheric boundary. This is estimated by the following metric~\citep{Prasad18}:
\begin{equation}
\label{eqChrom15}
E_n =\left(\sum_{i=1}^M |\textbf{$\emph{B}_{t,i}$}-\textbf{$\emph{b}_{t,i}$}|\times|\textbf{$\emph{B}_{t,i}$}|\right)/\left(\sum_{i=1}^M |\textbf{$\emph{B}_{t,i}$}|^2\right),
\end{equation}
where $M=N^2$ represents the total number of grid points on the transverse plane. To minimize the contribution from the weaker fields, the grid points are here weighted with respect to the strength of the observed transverse field; see~\citet{Hu10} for further details.

The extrapolated field $\textbf{\emph{B}}$ is a solution of an auxiliary higher-curl equation
\begin{equation}
\label{eqChrom16}
\mathbf{\nabla}\times\mathbf{\nabla}\times\mathbf{\nabla}\times\textbf{\emph{B}}+a_1 \mathbf{\nabla}\times\mathbf{\nabla}\times\textbf{\emph{B}}+b_1 \mathbf{\nabla}\times\textbf{\emph{B}}=0,
\end{equation}
which contains a second order derivative $(\mathbf{\nabla}\times\mathbf{\nabla}\times\textbf{\emph{B}})_z=-\mathbf{\nabla}^2 B_z$ at $z=0$ necessitating the requirement of vector magnetograms at two or more layers for evaluating the $\textbf{\emph{B}}$. To work with the available single layer vector magnetograms, an algorithm was devised by~\citet{Hu10} that involved additional iterations to successively correct the potential subfield $\textbf{$\emph{B}_{2}$}$. Starting with an initial guess, $\textbf{$\emph{B}_{2}$}=0$, the system is reduced to second order which allows for the determination of boundary conditions for $\textbf{$\emph{B}_{1}$}$ and $\textbf{$\emph{B}_{3}$}$ using the trial-and-error process described above. If the resulting minimum $E_n$ value is not satisfactory, then a corrector potential field to $\textbf{$\emph{B}_{2}$}$ is derived from the difference transverse field, i.e., $\textbf{$\emph{B}_{t}$}-\textbf{$\emph{b}_{t}$}$, and added to the previous $\textbf{$\emph{B}_{2}$}$ in anticipation of better agreement between the transverse fields as measured by $E_n$. This optimization procedure during which $E_n$ is iteratively minimized is automatic. In the present case, to minimize the computational cost, we ran the code for 1000 iterations during which we noted that $E_n$ asymptotically reached a value of 0.15. The algorithm relies on the implementation of fast calculations of the LFFFs including the potential field. Such extrapolations have been applied recently to model initial fields for flares and jets~\citep{Prasad18,Mitra18,Liu20}.

\begin{figure}[!http]
\centering
\includegraphics[width=0.48\textwidth]{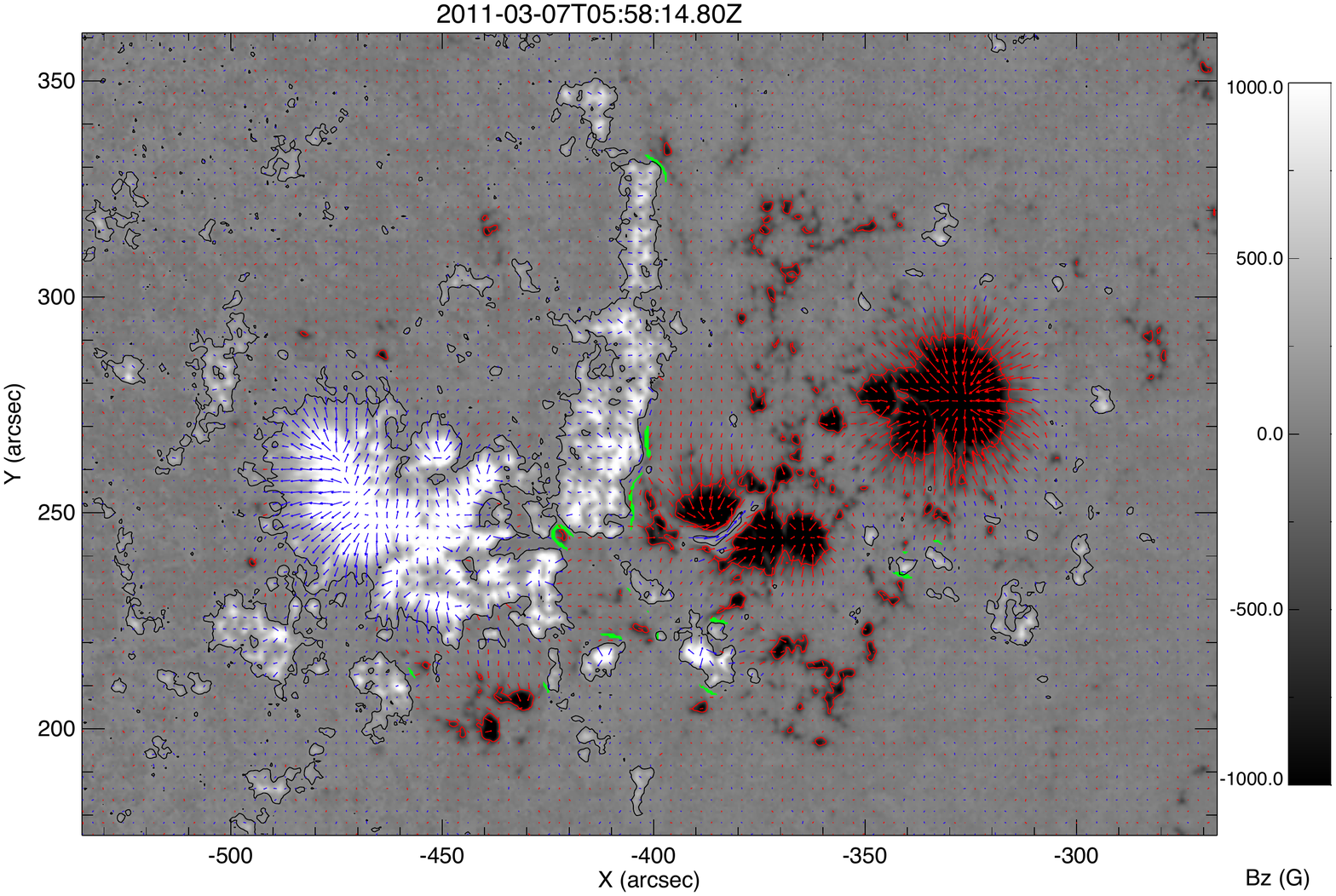}
\includegraphics[width=0.48\textwidth]{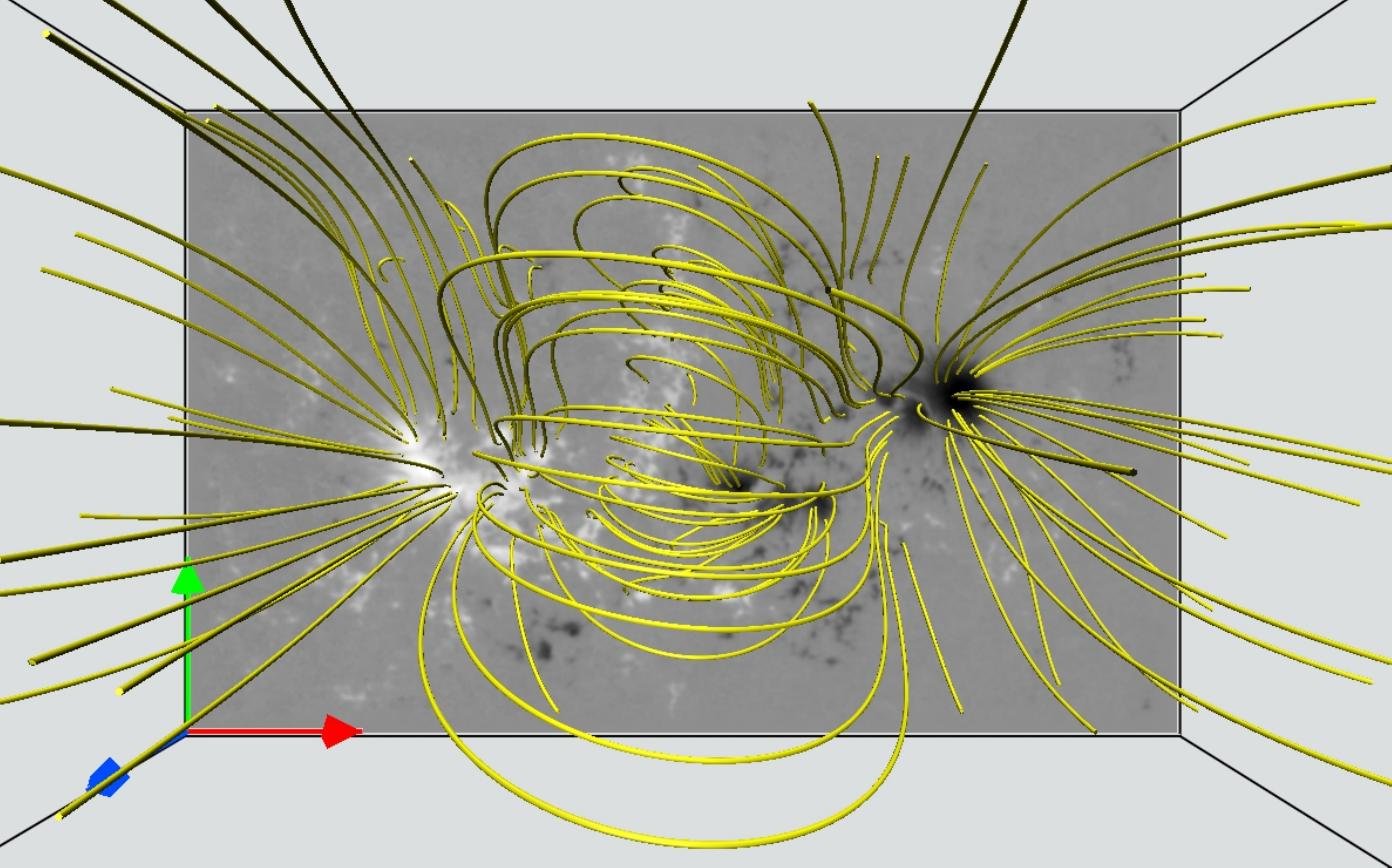}
\includegraphics[width=0.48\textwidth]{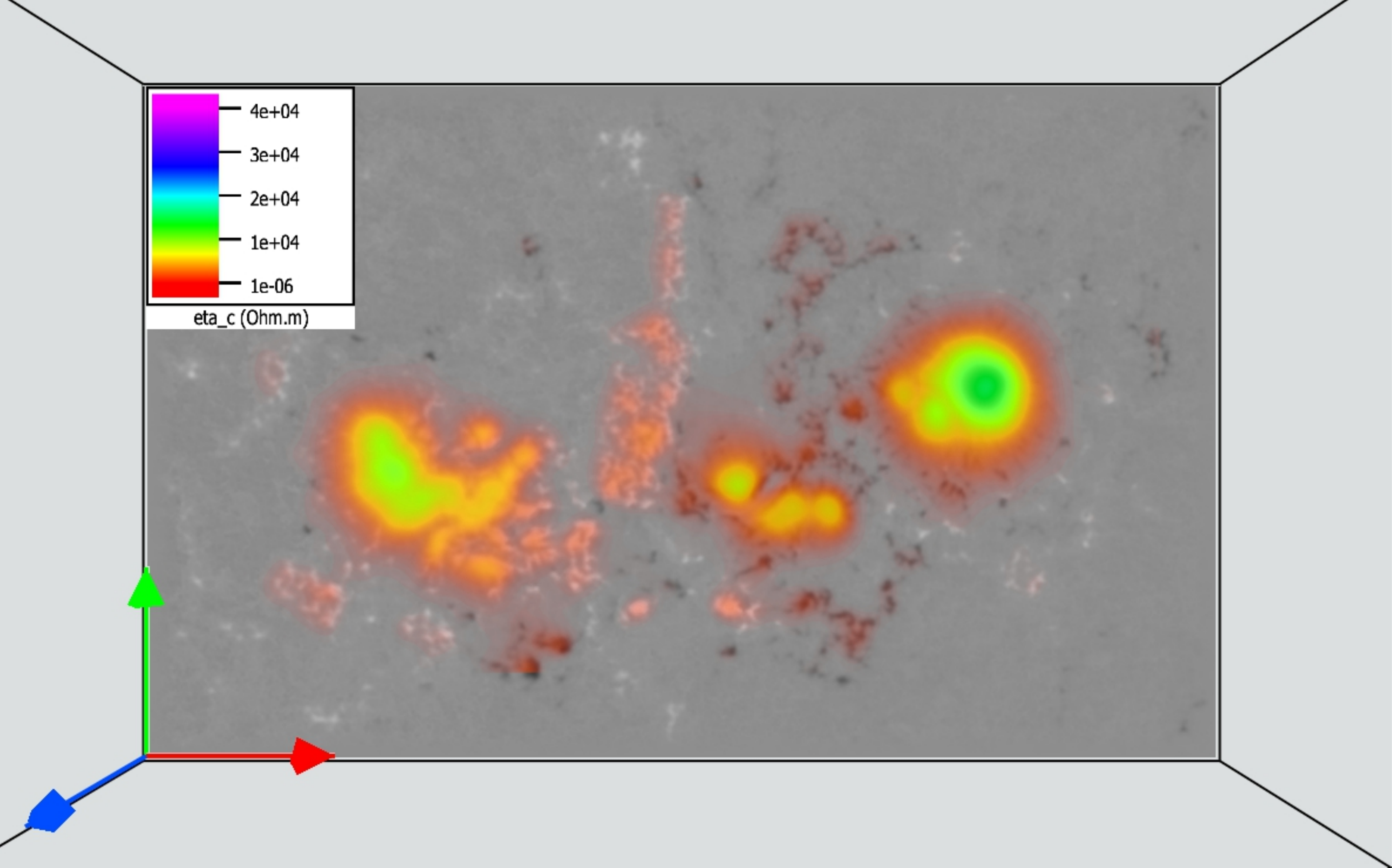}
\includegraphics[width=0.48\textwidth]{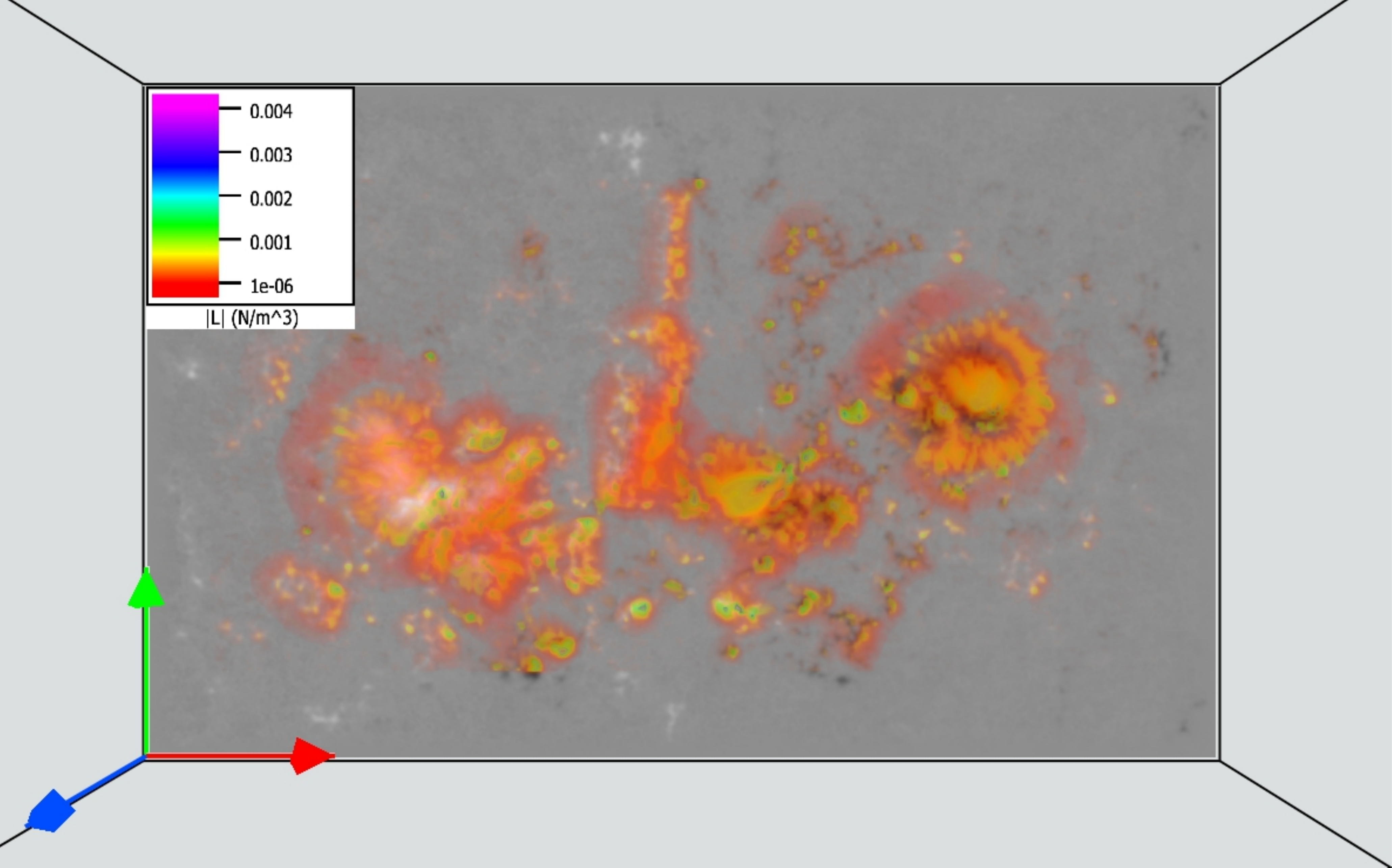}
\includegraphics[width=0.48\textwidth]{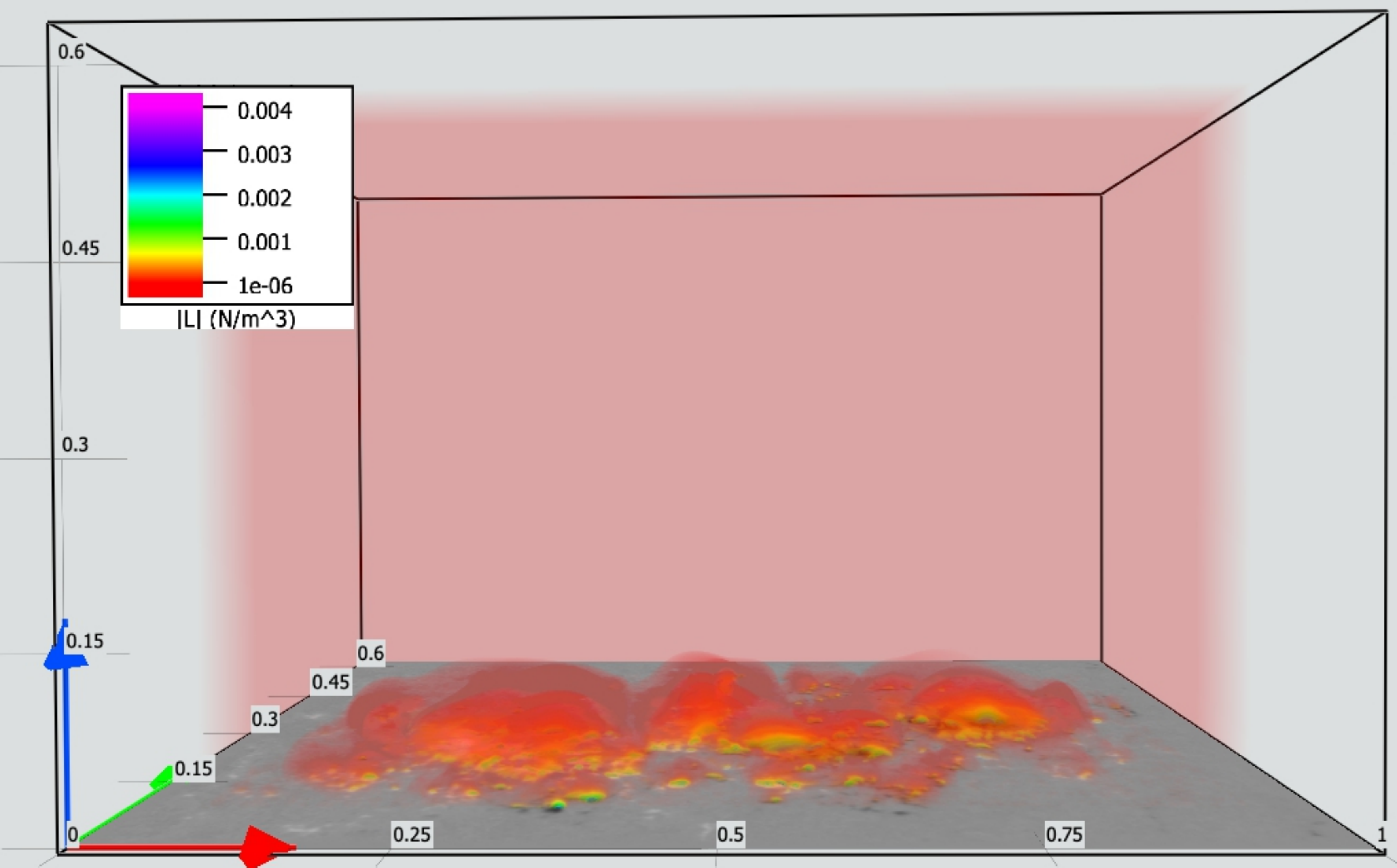}
\includegraphics[width=0.48\textwidth]{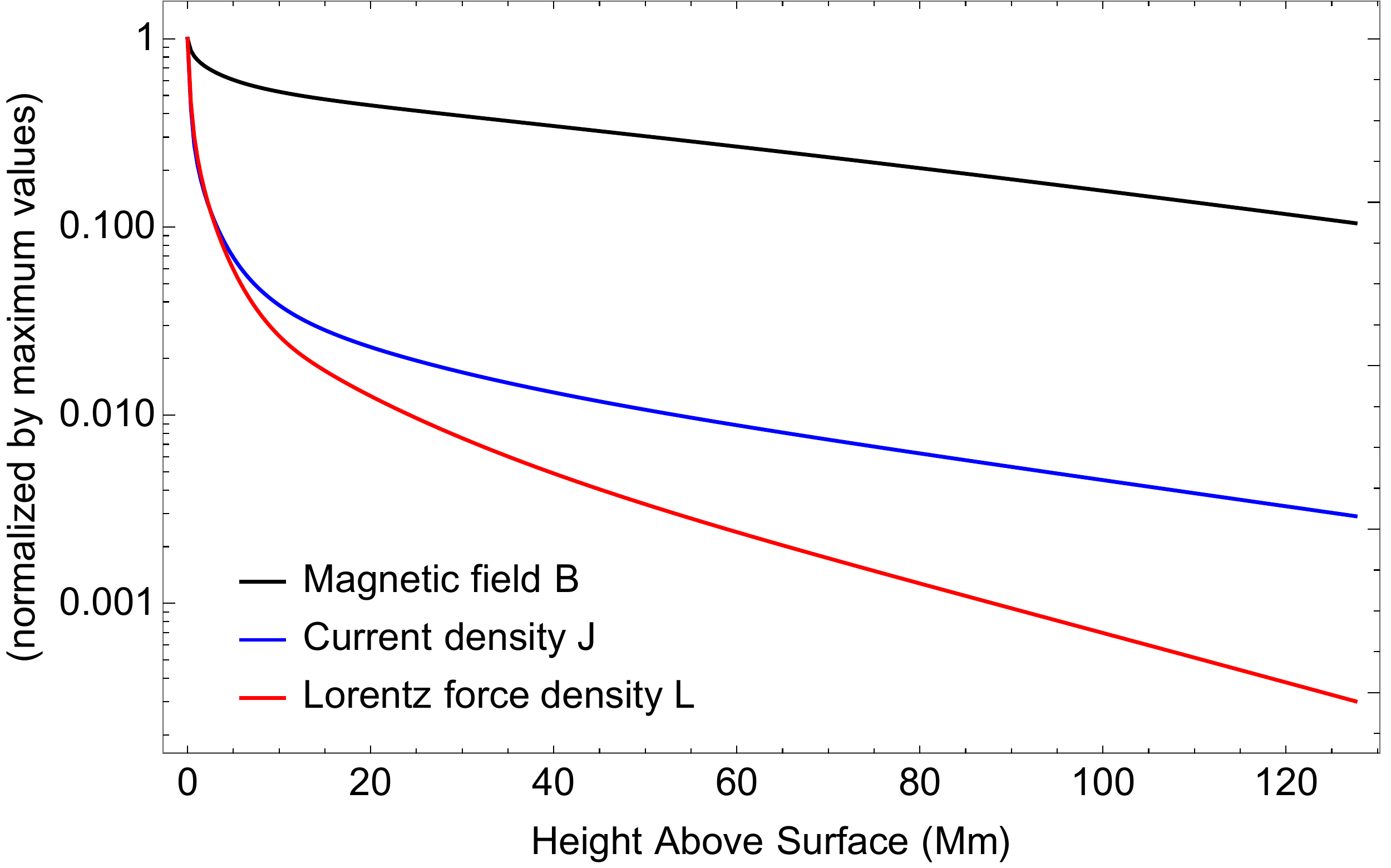}
\caption{(Top row) (left) HMI SHARP magnetogram showing the $B_{z}$ component with the PILs in green and transverse magnetic field vectors, and (right) NFFF magnetic fieldlines showing AR11166 at $t_{0}$; (middle row) Direct volume rendering of (left) $\eta_{C}$, and (right) Lorentz force density $\lvert$\textbf{$\emph{L}$}$\lvert$ between 1-2 Mm height; (bottom row) (left) Direct volume rendering of $\lvert$\textbf{$\emph{L}$}$\lvert$ with height showing the box dimensions, and (right) variations of $\lvert$\textbf{$\emph{B}$}$\lvert$, $\lvert$\textbf{$\emph{J}$}$\lvert$, and $\lvert$\textbf{$\emph{L}$}$\lvert$ with height. The values are averaged over the $xy$-plane at each height and normalized by their maxima. The heliographic coordinate axes are indicated in red, green, and blue arrows for $x$, $y$, and $z$ directions, respectively. The animation shows the evolution of AR11166 between $t_{0}$ and 2011-03-11T06:00:29 UT (hereafter called $t_{f}$) where Figure 2 (top left) is the first frame. The duration of the video is 4 s. (An animation of this figure is available.)}
\label{fig3}
\end{figure}

Figure~\ref{fig3} (top left) shows the HMI SHARP magnetogram data for the magnetic field component along the $z$-axis in the heliographic coordinate system, $B_{z}$, for AR11166 at 2011-03-07T06:00:29 UT (hereafter called $t_{0}$) with the polarity inversion lines (PILs) in green and transverse magnetic field vectors. Figure~\ref{fig3} (top right) shows the corresponding NFFF magnetic fieldlines. Figure~\ref{fig3} (middle left) shows that the Cowling resistivity is mostly important between 1-2 Mm height above the photosphere. The NFFF extrapolations for AR11166 were performed using the HMI vector magnetograms taken from the
``hmi.sharp\_cea\_720s" data series on a domain consisting of $560\times 352 \times 352$ ~pixels in the $x$, $y$ and $z$ directions, respectively. Since each pixel in HMI magnetogram corresponds to 0.5 arcsec, the horizontal extent of the box in $x$ corresponds to $\sim$200 Mm.

The NFFF extrapolations correctly capture the Lorentz force distribution, which is also principally concentrated around 1-2 Mm height (see Figure~\ref{fig3} middle right). The Lorentz force is then found to fall off sharply with height bringing the magnetic field close to a force-free state in the corona as shown in Figure~\ref{fig3} (bottom left and right). 

\section{Results \& Discussion} 
\label{res}

In this section, we will present results related to the variation of Cowling resistivity during the evolution of AR11166. We chose AR11166 for our analysis for the following reasons: (i) It has 3-5 days coverage within a meridional range of [-40$^{\circ}$,40$^{\circ}$] without any data gaps, and a flare occurs during this period and in this region; (ii) the early stage of the AR starts on the left limb and increases in complexity during its passage; and (iii) the AR is somewhat compact so that our NFFF extrapolations can be run at full resolution.

\subsection{Cowling resistivity variation during the evolution of AR11166}
\label{evol}

\begin{figure}[!http]
\centering
\includegraphics[width=0.49\textwidth]{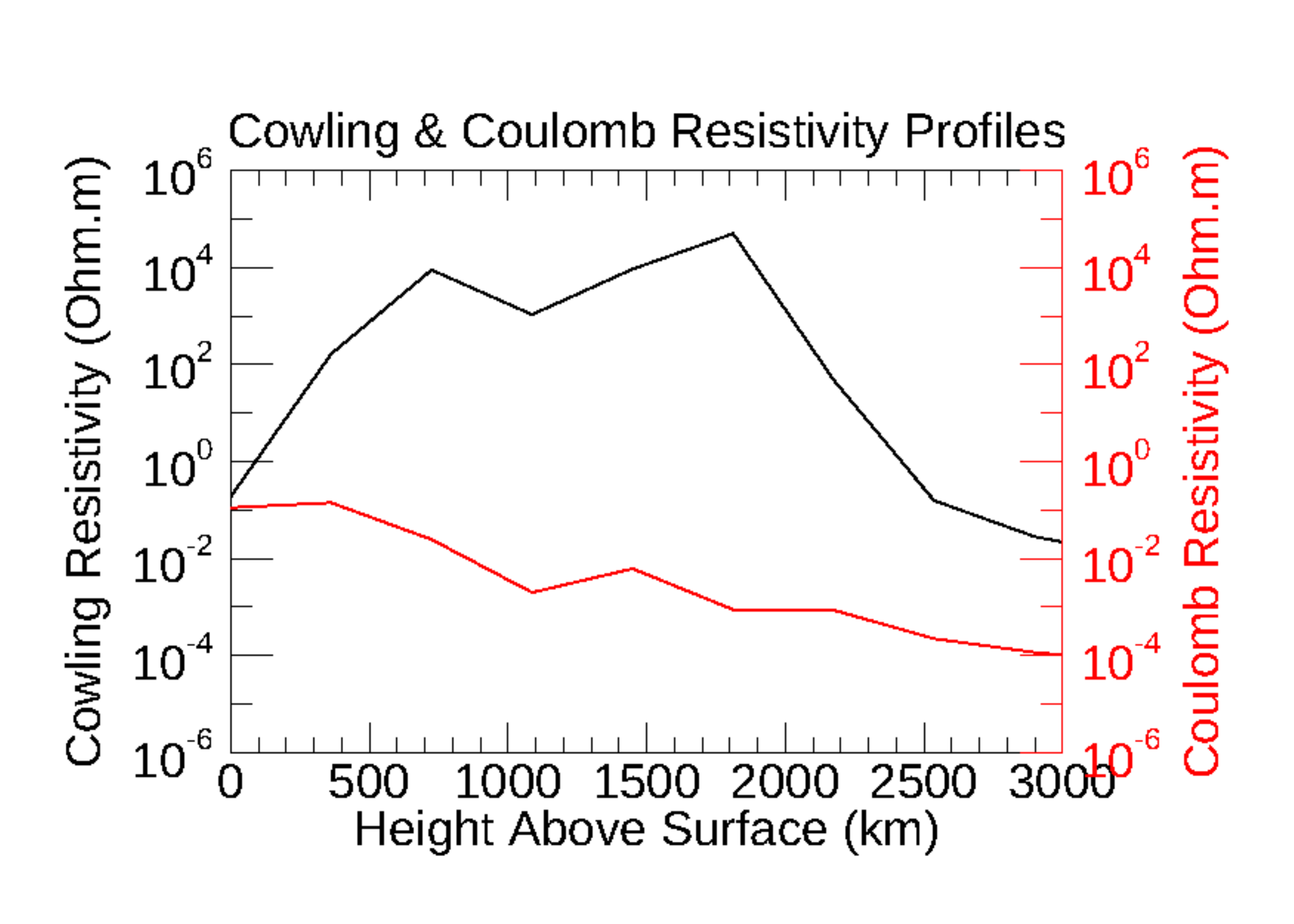}
\includegraphics[width=0.5\textwidth]{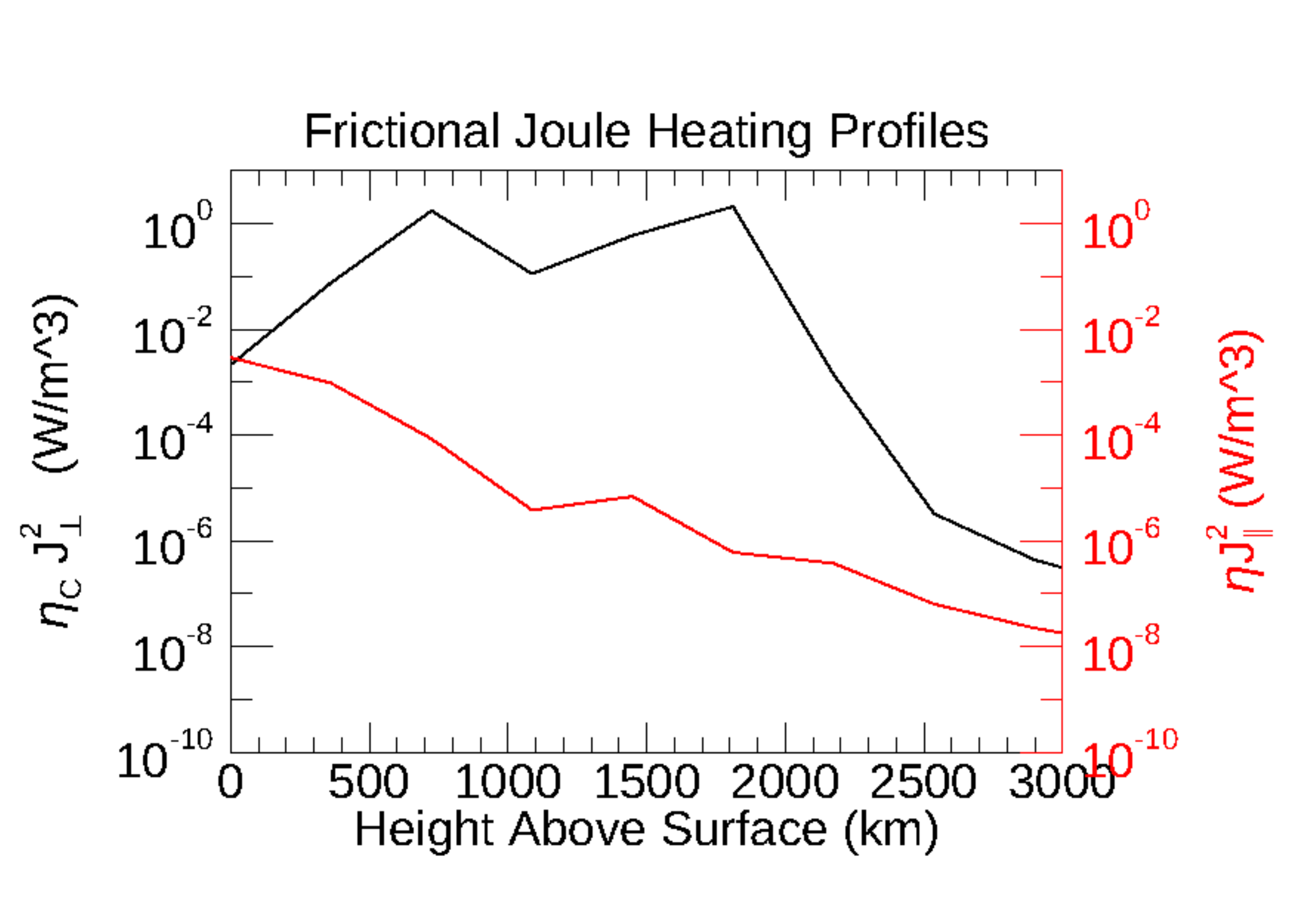}
\includegraphics[width=0.49\textwidth]{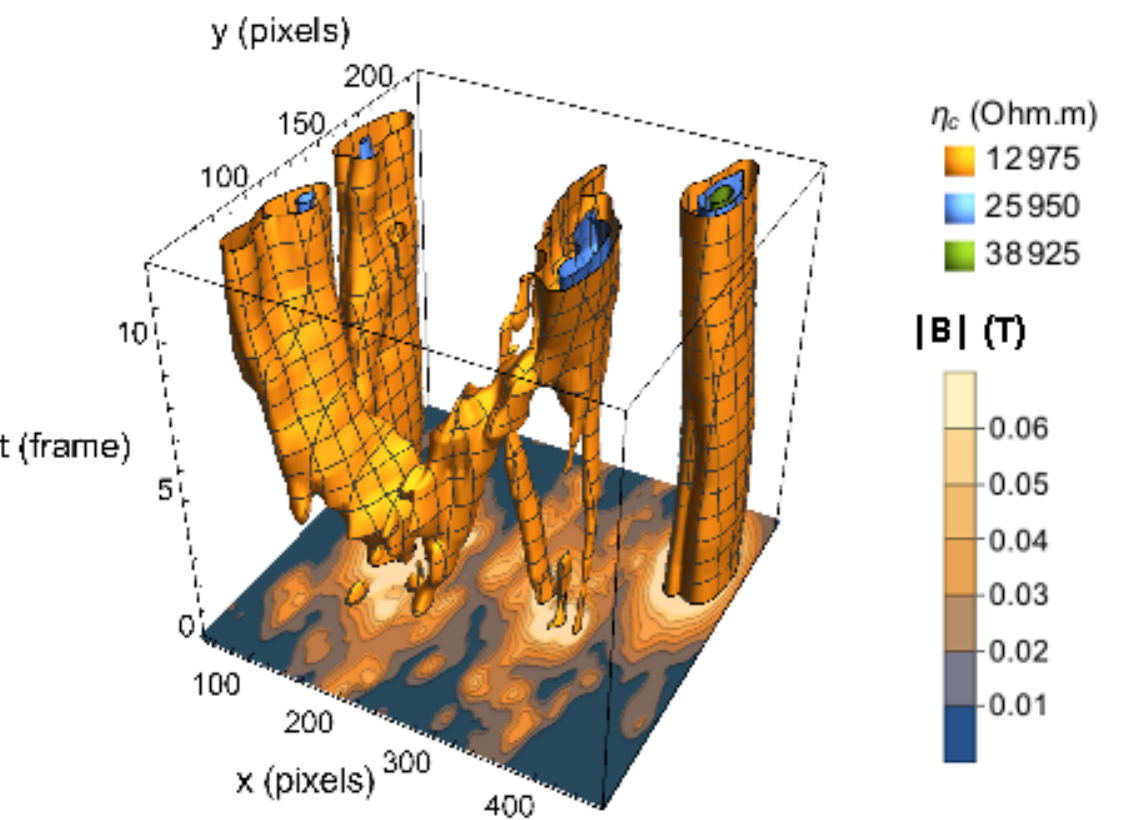}
\caption{Variations of (left) the maximum values of $\eta$ and $\eta_{C}$ profiles with height, and (right) the maximum values of frictional Joule heating profiles with height at $t_{0}$; (bottom) time-dependent variation of $\eta_{C}$ at $\sim$1.8 Mm height during the evolution of AR11166. The bottom boundary shows the variation of magnetic field strength at $\sim$1.8 Mm height at $t_{0}$.}
\label{fig4}
\end{figure}

We observe the evolution of AR11166 at 13 timesteps between $t_{0}$ and $t_{f}$ with a cadence of 8 hours. Figure~\ref{fig4} (left) shows the variations of the maximum values of Cowling and Coulomb resistivity profiles with height at $t_{0}$. The Cowling resistivity is orders of magnitude larger than the Coulomb resistivity in the chromosphere, specifically 6-8 orders of magnitude larger between 1-2 Mm. Figure~\ref{fig4} (right) presents the variation of the maximum values of the frictional Joule heating profiles with height in the chromosphere due to Cowling and Coulomb resistivities, showing that the chromospheric heating due to the dissipation of currents perpendicular to the magnetic field dominates the heating due to the dissipation of currents parallel to it. This figure demonstrates the significance of Cowling resistivity for chromospheric heating. Figure~\ref{fig4} (bottom) shows the time-dependent variation of Cowling resistivity at $\sim$1.8 Mm height during the evolution of AR11166. Since the Cowling resistivity distribution follows the AR structure quite well (see Figure~\ref{fig3} middle left) primarily due to its strong dependence on the magnetic field strength, its time variation shown in Figure~\ref{fig4} (bottom) can reveal how different structures on the AR evolve in time. Accordingly, the AR structures at the upper-right and upper-left do not change much as can be deduced from the vertical non-interacting contour structures whereas the other structures interact with each other above the PIL (see the animation corresponding to Figure~\ref{fig3} top left).

\subsection{The effect of Cowling resistivity on magnetic reconnection}
\label{magrecon}

Since the Cowling resistivity is orders of magnitude larger than the Coulomb resistivity in the chromosphere, it can in principle increase the magnetic reconnection rate significantly, and hence play a role in the flare formation, especially in a low-lying 3D null point configuration.

\begin{figure}[!http]
\centering
\includegraphics[width=0.48\textwidth]{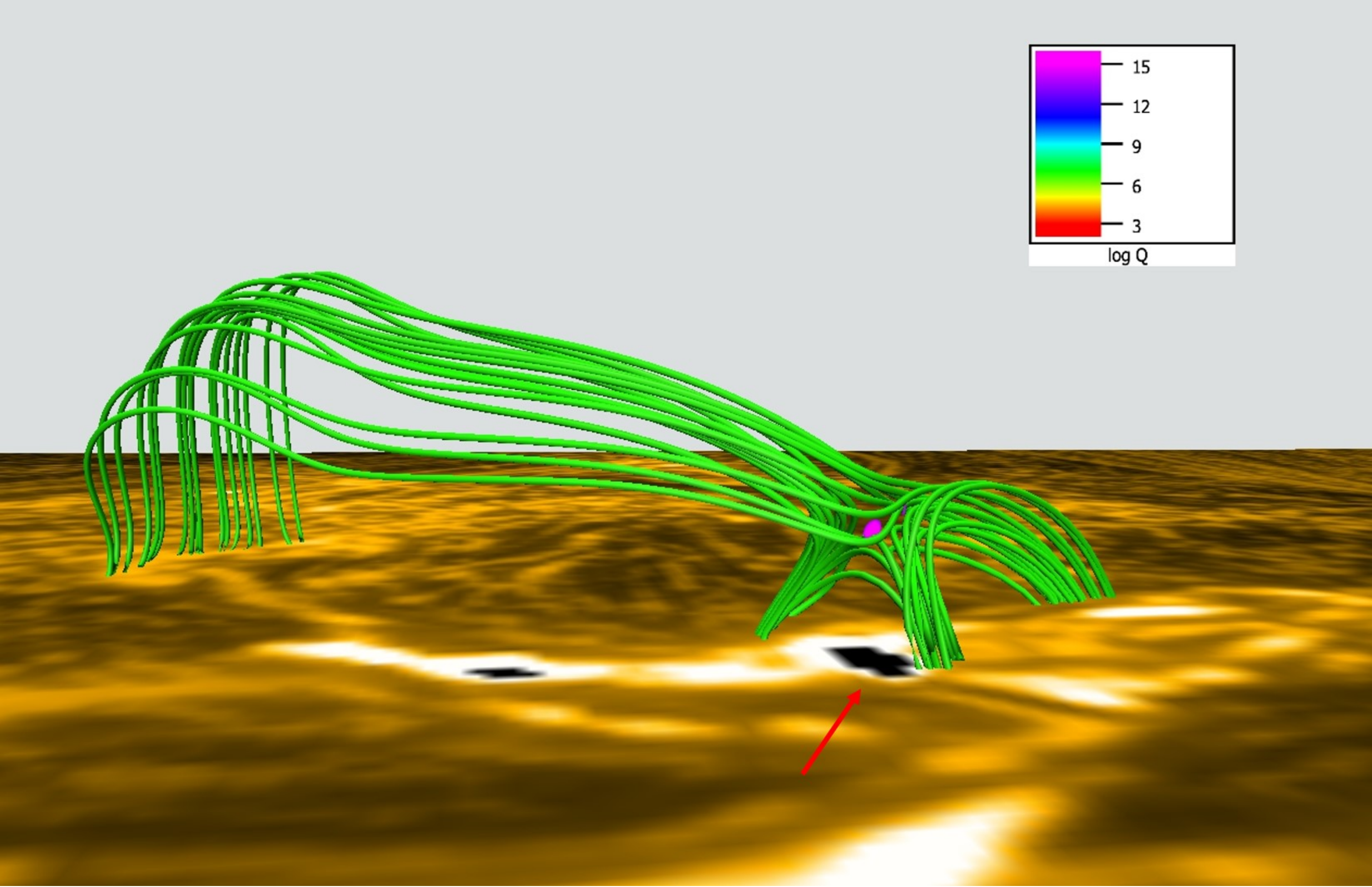}
\includegraphics[width=0.48\textwidth]{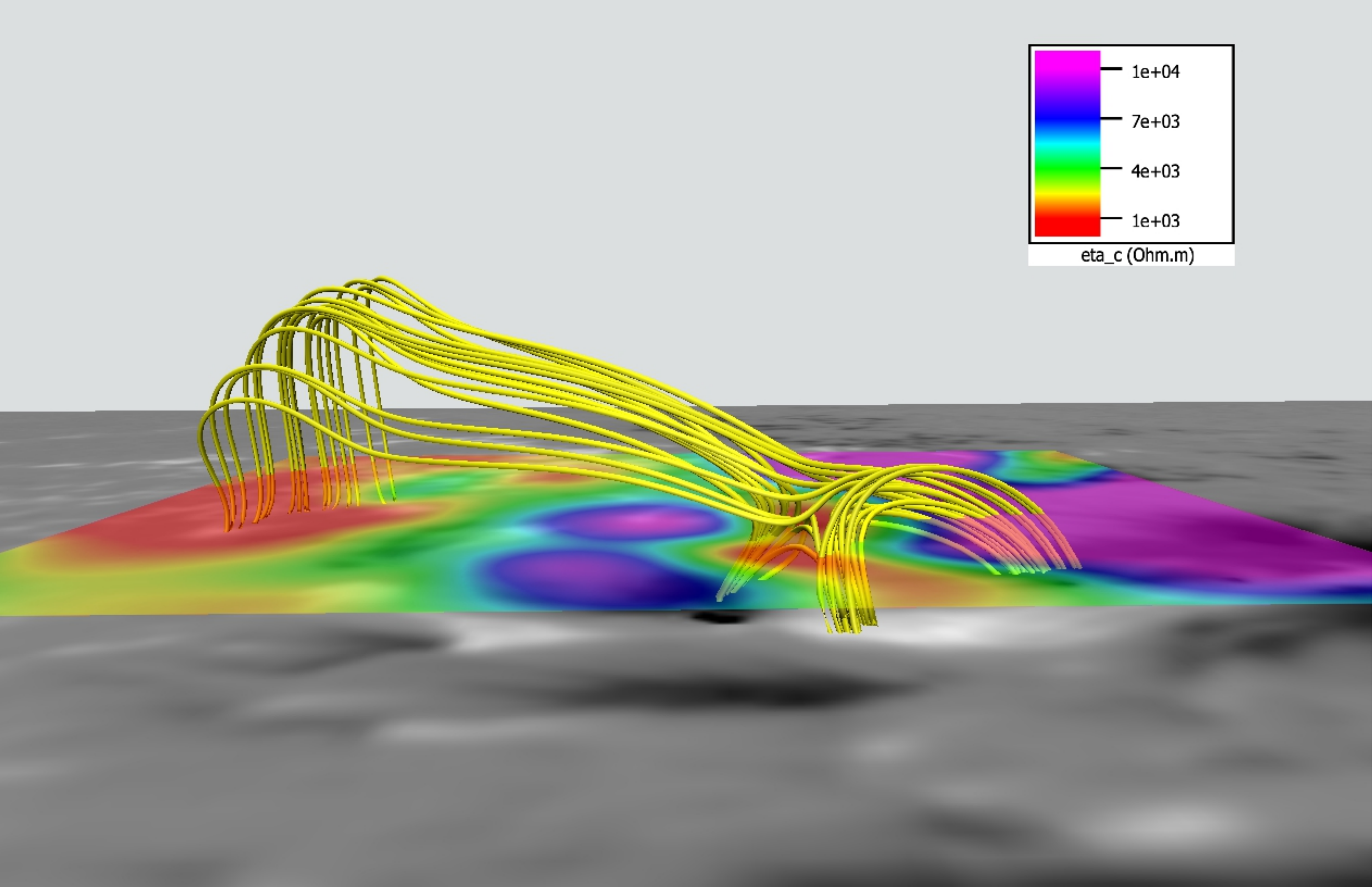}
\includegraphics[width=0.48\textwidth]{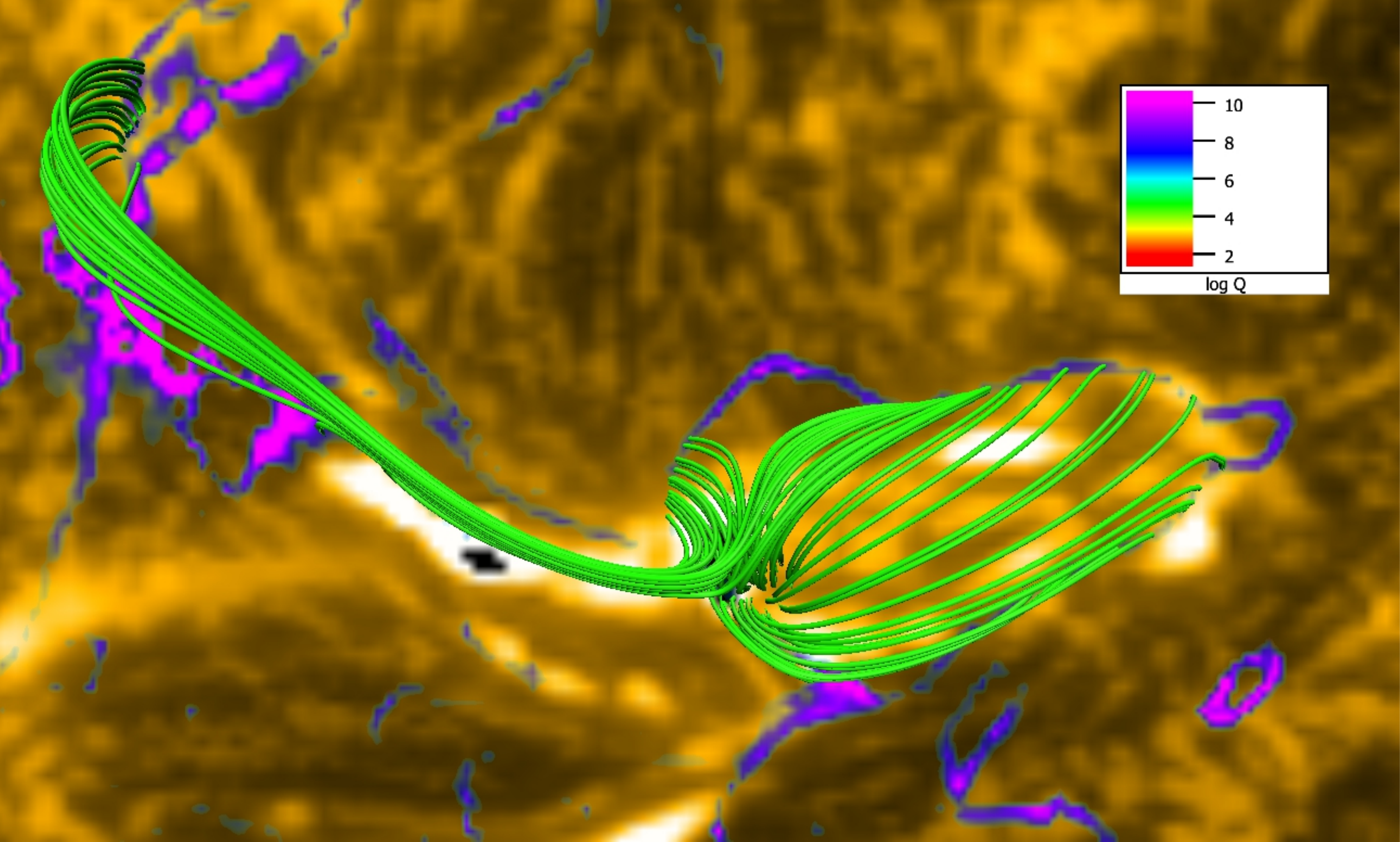}
\includegraphics[width=0.48\textwidth]{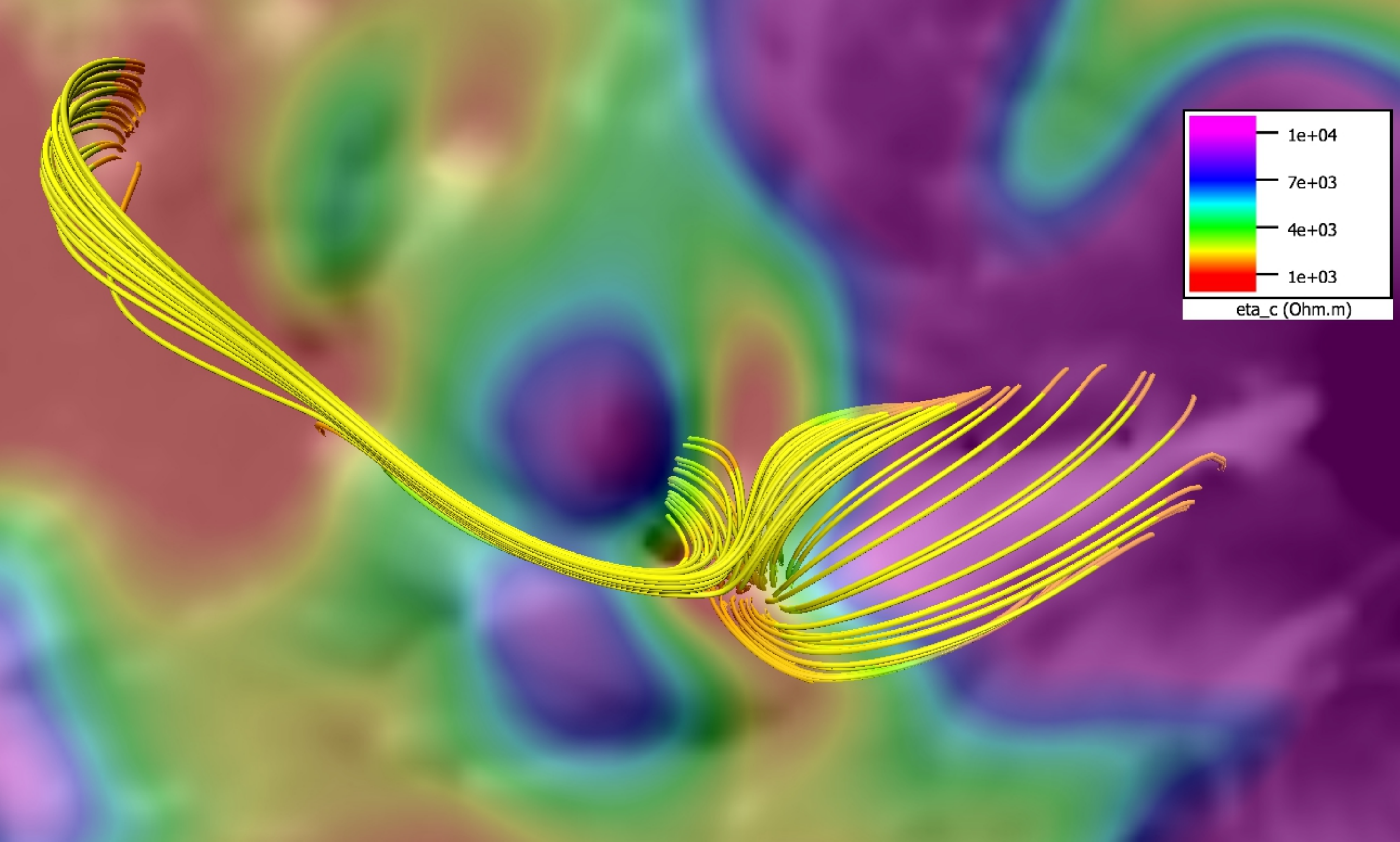}
\caption{(Left panel) Side and top views of a 3D null point with its corresponding spine-fan topology superimposed on an extreme-ultraviolet (EUV) channel 171 \AA~ image from the Atmospheric Imaging Assembly (AIA) \citep{Lemen12} onboard \emph{SDO} at 2011-03-10T14:23:36 UT corresponding to a C2.0 flare (indicated by the red arrow). The squashing factor (log Q) \citep{Liu16} contours are shown at the location of the 3D null point (top), which is at a height of $\sim$1.9 Mm, and at the bottom boundary (bottom). (Right panel) Side and top views of the magnetic field configuration superimposed on an HMI magnetogram showing AR11166 at 2011-03-10T14:24 UT and the $\eta_{C}$ distribution just below the null point. The C2.0 flare location is at (326,255) arcsec or (N15.34,W20.46) degrees.}
\label{fig5}
\end{figure}

In Figure~\ref{fig5}, we show such a 3D null point configuration at 2011-03-10T14:23:36 UT resulting in the C2.0 flare (see Figure~\ref{fig5} caption for details) emerging from AR11166 with its location being in a region where the Cowling resistivity is dominant.

According to~\citet{VL99}, the normalized magnetic reconnection rate (NRR) can be written as 
\begin{equation}
\label{eqChrom17}
\frac{v}{v_{A}}\approx\sqrt{\frac{\overline{\eta_{C}}}{v_{A}L}},
\end{equation}
where $\overline{\eta_{C}}=\eta_{C}/\mu_{0}$ and $\mu_{0}$ is the magnetic permeability in vacuum.

\begin{figure}[!http]
\centering
\includegraphics[width=0.45\textwidth]{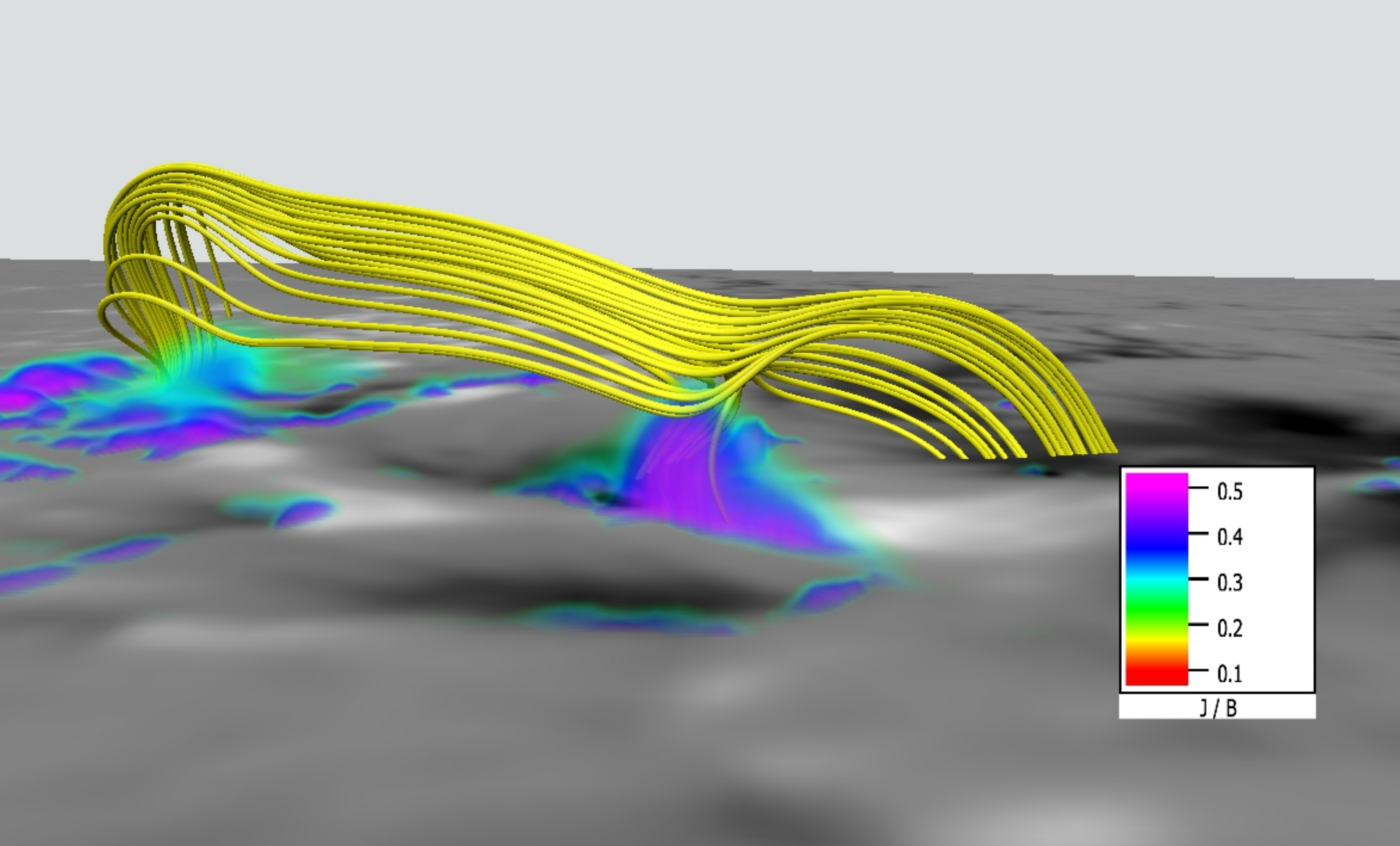}
\includegraphics[width=0.488\textwidth]{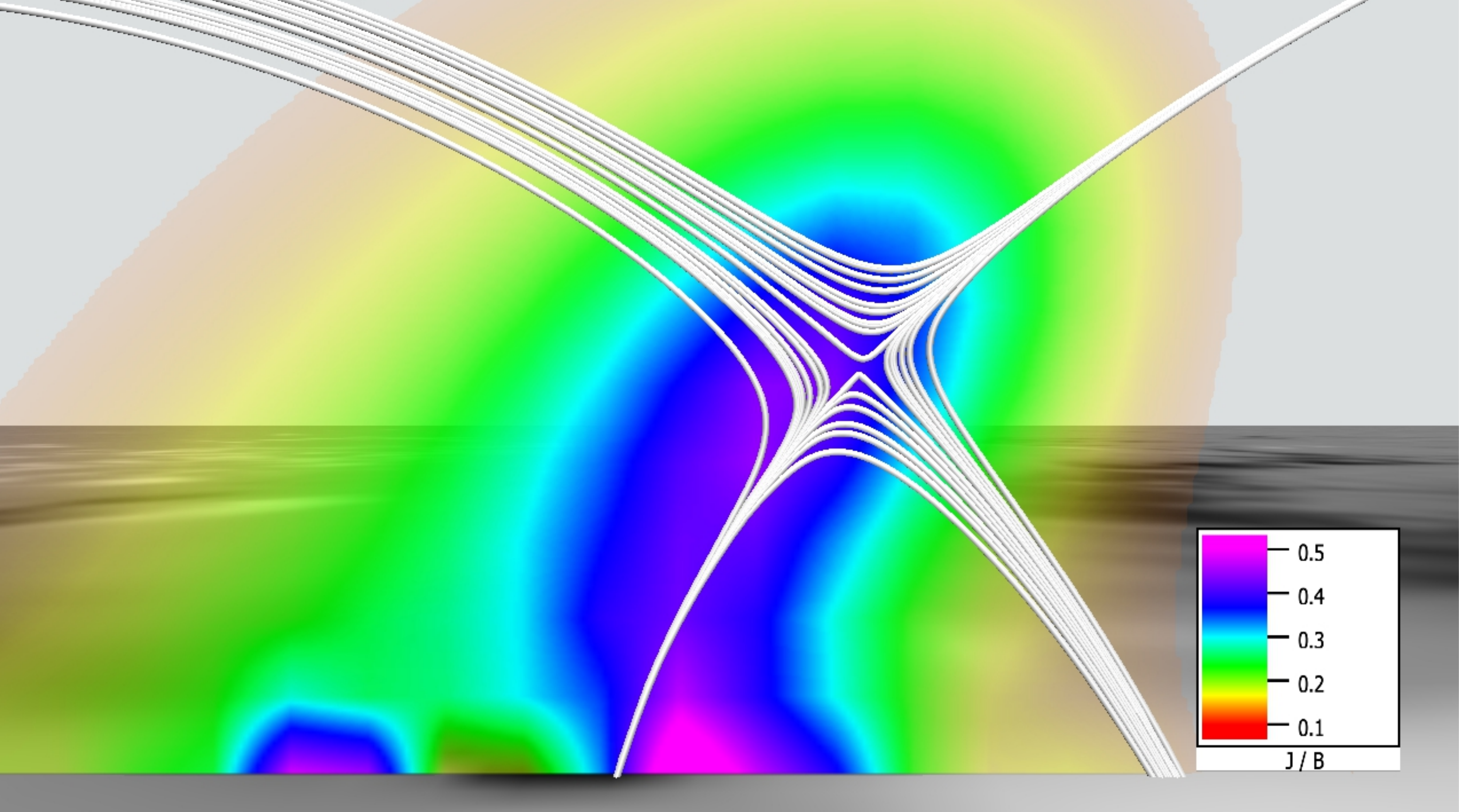}
\caption{(Left) Superimposed on HMI magnetogram showing AR11166 at 2011-03-10T14:24 UT is the volume distribution of $\lvert$\textbf{$\emph{J}$}$\lvert$/$\lvert$\textbf{$\emph{B}$}$\lvert$ close to the 3D null point just before the C2.0 flare; (right) a zoomed 2D view of the null point showing the $\lvert$\textbf{$\emph{J}$}$\lvert$/$\lvert$\textbf{$\emph{B}$}$\lvert$ contours in the magnetic reconnection current sheet. $\lvert$\textbf{$\emph{J}$}$\lvert$/$\lvert$\textbf{$\emph{B}$}$\lvert$ can be used as an indicator to estimate the width of the current sheet~\citep{Jiang16}.}
\label{fig6}
\end{figure}

In Figure~\ref{fig6}, the reconnection current sheet is shown using $\lvert$\textbf{$\emph{J}$}$\lvert$/$\lvert$\textbf{$\emph{B}$}$\lvert$ contours. Accordingly, $\lvert$\textbf{$\emph{J}$}$\lvert$/$\lvert$\textbf{$\emph{B}$}$\lvert$ $\sim$ 1/$L_{CS}$~\citep{Jiang16} where $L_{CS}$ is the width of the current sheet. The $\lvert$\textbf{$\emph{J}$}$\lvert$/$\lvert$\textbf{$\emph{B}$}$\lvert$ contour value in the current sheet is 0.5 which gives $L_{CS}=2$ pixels. The pixel size in our computation box for applying NFFF is 0.5 arcsec $\approx$ 362 km which is also equal to the half width of the current sheet, $L$, in Eq.~\ref{eqChrom17}. Taking $\overline{\eta_{C}}=1.5\times 10^9$ $\mathrm{m^2}$/s and the Alfv\'en wave speed $v_{A}=1563$ km/s in the vicinity of the 3D null point, NRR is found as 0.051. This value is in agreement with~\citet{Xue16} and the references therein.

According to the Sweet-Parker reconnection model, the characteristic half thickness of a current sheet can be written as
\begin{equation}
\label{eqChrom18}
l\sim\sqrt{\frac{\overline{\eta_{C}} L}{v_{A}}},
\end{equation}
which results in a current sheet thickness of 37 km compared to 25 m found for Coulomb resistivity.

Studies involving direct observational evidence of magnetic reconnection are relatively rare.~\citet{Xue16} estimates NRR as the Alfv\'enic Mach number of the inflow velocity similar to Eq.~\ref{eqChrom17}. They consider this as an upper limit for NRR since the outflow velocity may not generally reach the Alfv\'en velocity and the current sheet width obtained from the images is considered an upper limit for the actual width of the field reversal. Similarly, replacing the Coulomb resistivity with the Cowling resistivity as in Eq.~\ref{eqChrom17} gives an upper limit for NRR as the inflow velocity, $v$, in Eq.~\ref{eqChrom17} gives the maximum reconnection speed that is obtainable through ambipolar diffusion~\citep{VL99} or Cowling resistivity.

There are also numerical studies that simulate the magnetic reconnection in a partially-ionized chromosphere.~\citet{Leake12} solves a multi-fluid reacting hydrogen plasma model that takes ionization imbalance into account resulting in a reconnection rate which is almost independent of the Lundquist number. In this study, we follow the single-fluid approach due to~\citet{LA06} which implicitly assumes that the ions and neutrals are in ionization balance and follows the interactions between the ions and neutrals by the Cowling resistivity. Despite the differences between the models, our estimate for NRR of 0.051 is in agreement with the simulated and observed values in~\citet{Leake12} and~\citet{Xue16} and the references therein, respectively.

\section{Conclusions} 
\label{conc}

In this paper, we calculated the Cowling resistivity using the magnetic field obtained from NFFF extrapolation of photospheric vector magnetogram data, and density and temperature values from the Maltby-M model. We also discussed its effects in the weakly-ionized chromosphere during the evolution of an AR and on flare formation associated with magnetic reconnection in the chromosphere.

We analyzed the evolution of AR11166. The Cowling resistivity is found to be 6-8 orders of magnitude larger than the Coulomb resistivity between 1-2 Mm height in the chromosphere. It has a significant effect on the chromospheric heating via frictional Joule heating due to current dissipation perpendicular to the magnetic field. The time-dependent evolution of Cowling resistivity gives an indication about the AR evolution since it follows the AR structure quite closely due to its strong dependence on the magnetic field strength. 

We also analyzed the effect of Cowling resistivity on the formation of a C2.0 flare that emerged from AR11166. The Cowling resistivity can have an effect on flare formation for a low-lying 3D null point configuration that occurs at a height of $\sim$1.9 Mm where we have a significant Lorentz force distribution and the Cowling resistivity has its largest value. We obtain an NRR of 0.051 which is in agreement with~\citet{Leake12} and~\citet{Xue16} and the references therein, and a relatively thick current sheet with a thickness of 37 km in the chromosphere. We also find a good match between the AIA brightening with the log Q contours and the location of the null point inferred from the extrapolated magnetic field topology.

In future work, we will focus on analyzing the effects of Cowling resistivity for a flare using HMI SHARP vector magnetogram data with 12-minute cadence including observations before and after the flare. We also plan to utilize the Interface Region Imaging Spectrograph \citep[\emph{IRIS};][]{DePontieu14} data to determine the density and temperature structures of the corresponding AR with height in the chromosphere instead of the Maltby-M model.
 
We acknowledge support from the NSF EPSCoR RII-Track-1 Cooperative Agreement OIA-1655280. Any opinions, findings, and conclusions or recommendations expressed in this material are those of the author(s) and do not necessarily reflect the views of the National Science Foundation. M.S.Y. and N.P. acknowledge partial support from NASA LWS grant 80NSSC19K0075. A.P. and Q.H. acknowledge partial support from NASA grant 80NSSC17K0016 and NSF award AGS-1650854.  

The HMI and AIA data have been used courtesy of NASA/\textit{SDO}, and HMI and AIA science teams.

\end{document}